\documentclass[prb,aps,twocolumn,draft,tightenlines,%
floatfix,superscriptaddress]{revtex4}
\usepackage{amssymb}
\usepackage{amsmath}
\usepackage{colordvi}

\begin{document}

\title{Spin dynamics in semiconductor nanostructures\footnote{An invited 
talk to be presented at nanomeeting-2007 (http://www.nanomeeting.org/). To be published in
Physics, Chemistry and Application on Nanostructures, 2007 (World Scientific,
Singapore, 2007).}}

\author{M. W. Wu}
\thanks{Author to whom correspondence should be addressed.}
\email{mwwu@ustc.edu.cn.}
\affiliation{Hefei National Laboratory for Physical Sciences at
  Microscale, University of Science and Technology of China, Hefei,
  Anhui, 230026, China}
\affiliation{Department of Physics,
University of Science and Technology of China, Hefei,
  Anhui, 230026, China}
\author{M. Q. Weng}
\author{J. L. Cheng}
\affiliation{Department of Physics,
University of Science and Technology of China, Hefei,
  Anhui, 230026, China}

\date{\today}
\begin{abstract}
 We review our theoretical investigation on the spin
 relaxation/dephasing in spin precession and spin diffusion/transport
 in semiconductor nanostructures based on the kinetic spin Bloch
 equation approach. 
\end{abstract}

\maketitle

\section{Introduction}

Much attention has been devoted to the electron spin dynamics in
semiconductors for the past three decades.\cite{meier, prinz} Especially,
recent experiments have shown extremely long spin lifetime (up to
hundreds of nanoseconds) in -type bulk Zinc-blende semiconductors
(such as GaAs).\cite{kikkawa, dzhioev, murdin} Moreover, a lot more investigations have been
performed on various low dimensional systems.\cite{damen} The spin
diffusion/transport  has also been studied experimentally and very
long spin injection length are reported.\cite{ohno} These findings show the
great potential for using the spin degree of freedom in place of, or
in addition to, the charge degree of freedom for device application
such as qubits and spin transistors. A thorough understanding of the
spin relaxation/dephasing (R/D) in the spin precession and spin
diffusion/transport is essential  for such application.

It is understood that the D'ayakonov-Perel' (DP) mechanism is the
leading spin R/D mechanism in $n$-type Zinc-blende semiconductors.\cite{
yakonov} Many theoretical works have been carried out to study the spin
relaxation time in various systems\cite{meier, lau} based on the
single-particle formula\cite{meier}
\begin{equation}
  \label{SDT_elec}
  \frac {1}{\tau} =\frac{ \int^{\infty}_{0}dE_k
    (f_{k1/2}-f_{k-1/2})\tau_p(k)\overline{{\bf h}^2({\bf k})} }
{2\int^{\infty}_{0}dE_k(f_{k1/2}-f_{k-1/2})}\ .
\end{equation}
Here $\tau_p(k)$ is the momentum relaxation time which is due to the
electron-phonon and electron-impurity scattering. $f_{{\bf k} \sigma}$
stand for the
electron distribution functions of spin $\sigma$. ${\bf h}({\bf k})$ is the DP
term which serves
as an effective magnetic field and is composed of the Dresselhaus
term\cite{dresselhaus} due to the bulk inversion asymmetry (BIA) and
the Rashba term\cite{bychkov}
due to the structure inversion asymmetry (SIA). $\overline{{\bf
    h}^2({\bf k})}$ denotes the
average of ${\bf h}^2({\bf k})$ over all directions of ${\bf k}$. In
GaAs quantum well (QW), the Dresselhaus term is the leading term and
${\bf h}({\bf k})$ has the form:
\begin{eqnarray}
\label{eq2}
   h_x({\bf k}) &=& \gamma k_x(k_y^2-\langle k_z^2\rangle)\ ,
   \nonumber \\
   h_y({\bf k}) &=& \gamma k_y(\langle k_z^2\rangle-k_x^2)\ ,
   \nonumber \\
   h_z({\bf k}) &=& 0 \ ,
\end{eqnarray}
in which $\langle k_z^2\rangle$ represents the average of the operator
$(\partial/\partial z)^2$ over the electronic state of the lowest
 subband. $\gamma$ is the Dresselhaus spin-orbit
parameter.\cite{meier, aronov} For InAs, the Rashba term is more 
important and ${\bf h}({\bf k})$ is given by
$h_x({\bf k})=\alpha k_x$, $h_y({\bf k})=\alpha k_y$ and $h_z({\bf k})=0$,
in which the Rashba coefficient $\alpha$
is  proportional to the interface electric
field $E_z$ along the growth direction: $\alpha=\alpha_0E_z$,
with the coefficient $\alpha_0$ being inversely proportional to the
energy gap and the effective mass.\cite{lommer}
Equation (\ref{SDT_elec}) is valid only when 
$|{\bf h}|\tau_p\ll1$, {\em i.e.}, the strong scattering regime,
 and the scattering is elastic. It also cannot be applied to 
system far away from equilibrium, such as system with
large spin polarization and/or with a strong in-plane electric field.
Moreover, the Coulomb scattering has long been 
neglected as it does not contribute to the momentum relaxation
directly.

It was shown recently by Wu {\em et al}. from a full microscopic
kinetic-spin-Bloch-equation (KSBE) approach that the single-particle
approach is inadequate in accounting for the spin R/D both in the time
domain\cite{wu, wu2, wu3, weng, weng2} and in the space
domain.\cite{weng3, weng4, weng5, jiang} The momentum
dependence of the effective magnetic field (the DP term) and the
momentum dependence of the spin diffusion rate along the spacial
gradient\cite{weng3} or even the random spin-orbit coupling (SOC)\cite{ya} all
serve as inhomogeneous broadenings.\cite{wu2, wu3} It was pointed out that in
the presence of inhomogeneous broadening, any scattering, including
the carrier-carrier Coulomb scattering, can cause an irreversible spin
R/D.\cite{wu2, wu3} Moreover, besides the spin R/D channel the scattering
provides, it also gives rise to the counter effect to the
inhomogeneous broadening. The scattering tends to drive carriers to a
more homogeneous state and therefore suppresses the inhomogeneous
broadening. Finally, this approach is valid in both strong and weak
scattering limits and can be used to study systems far away from the
equilibrium.

In this paper, we review the KSBE approach in various nanostructures
and under different conditions. The paper is organized as follows: In
Sec. 2 we set up the KSBEs. In Sec. 3 we review the results of the
spin R/D in the time domain. The results of the spin R/D in the spin
transport/diffusion are reviewed in Sec. 4. We conclude in Sec. 5.

\section{Model and KSBEs}

We start our investigation from an $n$-type zinc-blende semiconductor QW
with the growth direction along the $z$-axis. The well width is assumed
to be small enough so that only the lowest subband is
relevant. Sometimes a moderate magnetic field {\bf B} is applied along the
$x$-axis (in the Voigt configuration). 

By using the nonequilibrium Green function method with gradient
expression as well as the generalized Kadanoff-Baym Ansatz,\cite{haug} we
construct the KSBEs as follows:
\begin{eqnarray}
\dot \rho_{\mathbf{k}}(\mathbf{r}, t) &=&
\left.\dot\rho_{\mathbf k}(\mathbf{r}, t)
\right|_{\mathtt{dr}}
+ \left. \dot \rho_{\mathbf k}(\mathbf{r}, t)
\right|_{\mathtt{dif}}\nonumber\\
&&+ \left. \dot \rho_{\mathbf k}(\mathbf{r}, t)
\right|_{\mathtt{coh}}
+ \left. \dot\rho_{\mathbf k}(\mathbf{r},t)\right|_{\mathtt{scat}}\ .
\label{KSBE}
\end{eqnarray}
Here $\rho_{\mathbf k}(\mathbf
r, t)=\begin{pmatrix}f_{\mathbf k\uparrow} & \rho_{\mathbf
    k\uparrow\downarrow}\\
  \rho_{\mathbf k\downarrow\uparrow} & f_{\mathbf
    k\downarrow}\end{pmatrix}$ are the density matrices of electrons
with momentum ${\bf k}$ at position ${\bf r}=(x,y)$ and time $t$.
The off-diagonal elements $\rho_{\mathbf
  k\uparrow\downarrow}=\rho_{\mathbf k\downarrow\uparrow}^{\ast}$ represent
the correlations between the spin-up and -down states.
$\left.\dot\rho_{\mathbf k}(\mathbf{r}, t)\right|_{\mathtt{dr}}=
\{\mathbf{\nabla}_{\mathbf{r}}
\overline{\varepsilon}_{\mathbf{k}}(\mathbf{r}, t),
\mathbf{\nabla}_{\mathbf{k}}\rho_{\mathbf k}(\mathbf{r}, t)\}/2$
are the driving terms from the external electric field.
Here $\overline{\varepsilon}_{\mathbf{k}} (\mathbf{r}, t) = \mathbf{
k}^2/2m^{\ast} +[g\mu_B{\mathbf B}+{\mathbf h}(\mathbf
k)]\cdot \mbox{\boldmath$\sigma$\unboldmath}/2-e\Psi(\mathbf
r)+{\cal E}_{\mathtt{HF}}({\bf r},t)$
with ${\cal E}_{\mathtt{HF}}({\bf r},t)=-\sum_{\mathbf  q}
V_{\mathbf q}\rho_{\mathbf{k-q}}
(\mathbf r,t)$ representing the Hartree-Fock (HF) term.
$m^{\ast}$ is the effective mass. $\Psi (\mathbf r)$ is determined
from the Poisson equation. The bracket $\{A, B\}=AB+BA$ is the
anti-commutator. $\left.\dot\rho_{\mathbf k}(\mathbf{r}, t)
\right|_{\mathtt{dif}}=-\{\mathbf{\nabla}_{\mathbf{k}}\overline
{\varepsilon}_{\mathbf{k}}(\mathbf{r},t),\mathbf{\nabla}_
{\mathbf{r}}\rho_{\mathbf k}(\mathbf{r},t)\}/2$ represent the
diffusion terms.
The coherent terms in Eq.\ (\ref{KSBE}) are $
\dot\rho_{\bf k}|_{\mathtt{coh}} =
i[(g\mu_B{\bf B}+{\bf h}({\bf k}))\cdot\mbox{\boldmath$\sigma$\unboldmath}
/2+{\cal E}_{\mathtt{HF}},\ \rho_{\mathbf{k}}]$.

The scattering terms are different depending on different 
statistics.\cite{cheng} One is the collinear statistics where the
equilibrium state is taken as the Fermi distribution of electrons in the
conduction band without SOC term (DP term). Therefore the
energy spectrum is $\varepsilon_k=k^2/2m^\ast$ and
the eigenstates of spin are the eigenstates of
$\sigma_z$, {\em i.e.} $\chi_\uparrow=(1,0)^T$
and $\chi_\downarrow=(0,1)^T$. The other is the helix statistics 
where the equilibrium state is taken to be the Fermi distribution of
electrons in the conduction band 
with the SOC. The energy spectrum is then being $
\varepsilon_{{\bf k},\xi}=
k^2/2m^\ast+\xi|{\bf h}({\bf k})|$ 
 with $\xi=\pm 1$ for the two spin branches.
The eigenfunctions of spin are therefore
$|\xi\rangle=
[\chi_{\uparrow} + \xi {\tilde h}(\mathbf{k})  \chi_{\downarrow}/
{|{\bf h}(\mathbf{k})|}]/ \sqrt{2}$
with ${\tilde h}({\bf k})
=h_x({\bf k})+ih_y({\bf k})$. In principle, the helix statistics is
the correct statistics. Nevertheless, as the SOC is very small in
semiconductors compared to the Fermi energy, the SOC in the energy
spectrum can be neglected.\cite{cheng} This also facilitates an accurate
numerical calculation of the scattering.

The spin R/D times can be determined from the time evolution of the
density matrix $\rho_{\bf k}$ by
numerically solving the KSBEs with all the scattering explicitly 
included. The spin relaxation time $T_1$ 
is determined from the slope of the envelope of 
$\Delta N=\sum_{\bf k}(f_{{\bf k},\uparrow}-f_{{\bf k},\downarrow})$.
The irreversible spin dephasing time $T_2$ is associated with the
incoherently summed spin coherence\cite{wu}
$\rho= \sum_{\bf k}|\rho_{{\bf k}\uparrow\downarrow}(t)|$ 
whereas the ensemble spin dephasing time 
$T^{\ast}_2$ is  defined from the slope of the
envelope of the coherently summed spin coherence
$\rho^{\prime}= |\sum_{\bf k}\rho_{{\bf k}\uparrow\downarrow}(t)|$.
Similarly for spin diffusion/transport, the spin diffusion length can be
determined from the spacial evolution of $\Delta N$ in the steady 
state.\cite{weng3}

\section{Spin R/D}

In this section we review our results of spin R/D in semiconductor
nanostructures based on SKBE approach. It was shown that any spin
conserving scattering, including the electron-electron Coulomb
scattering, can cause an irreversible spin R/D in the presence of the
inhomogeneous broadening.\cite{wu2,wu3,weng,weng2} The energy dependence of the g
factor\cite{wu2} and the momentum dependence of the DP
term\cite{wu3,weng,weng2} all
serve as the inhomogeneous broadening. In quantum wire, the
inhomogeneous broadening can be quantified by the standard deviation
of the Larmor frequencies which well corresponds to the resulting spin
dephasing time.\cite{cheng2} It is further shown in GaAs (110) QWs that in the
presence of a magnetic field in the Voigt configuration, there is
inhomogeneous broadening from the DP term and the spin R/D time is
finite, although much longer than that in (001) QWs.\cite{wu4} A thorough
many-body investigation on $n$-type GaAs (001) QWs is performed with all
the scattering explicitly included\cite{weng} when the temperature is higher
than 120 K. It is shown that in QW with small well width, the spin R/D
time increases with the temperature in stead of decreases as predicted
from the single-particle approach when the electron density is in the
order of $10^{11}$ cm$^{-2}$. This temperature dependence is in good agreement with the
experimental result by Malinowski {\em et al.}\cite{damen} For larger well width,
the situation may become different. Weng and Wu calculated the spin
R/D for larger well widths by including the multi-subband effect.\cite{weng6} It is shown that for small/large well width so that the
linear/cubic term in Eq. (\ref{eq2}) is dominant, the spin R/D time
increases/decreases with the temperature. This is because with the
increase of temperature, both the inhomogeneous broadening and the
scattering get enhanced. The relative importance of these two
competing effects is different when the linear/cubic term is
dominant.\cite{weng6} As the Coulomb scattering is included to orders of all the
bubble diagrams, one is able to calculate the spin R/D far away from
the equilibrium. It is shown in Ref.\ \onlinecite{weng} that the spin R/D time increases
dramatically with the spin polarization. This is discovered due to the
lowest order (HF) contribution of the Coulomb interaction which
appears in the coherent term of the KSBEs. With high spin
polarization, the HF term serves as an effective magnetic field along
the $z$ axis which blocks the spin precession. It is further shown in
Ref.\ \onlinecite{weng2} that after some tricks in the numerical scheme, one is able to
calculate the spin R/D in the presence of a high in-plane electric
field so that the system is in the hot-electron regime. It is shown
that in the presence of an in-plane electric field, electron spin can
precess in the absence of any magnetic field at high temperature. This
is understood that the in-plane electric field induces a
center-of-mass shift of the momentum which gives rise to an effective
magnetic field proportional to the electric field.\cite{weng2} The effect of
strain on the spin R/D is also discussed and it is shown that one can
effectively manipulate the spin R/D time by strain.\cite{jiang2} Cheng and Wu
further discussed the spin R/D under identical Dresselhaus and Rashba
terms.\cite{cheng} A finite spin R/D time is obtained due to the cubic term
in Eq. (2). Very recently, Zhou {\em et al}.\cite{zhou} extended the theory to the
regime of very low temperature by figuring out ways to deal with the
electron-AC phonon scattering numerically. An excellent agreement with
experiment by Ohno {\em et al}.\cite{damen} is obtained from $20$\ K to $300$\ K without
any fitting parameter. The Dresselhaus coupling parameter $\gamma$ is found
to be well represented by  $\gamma =
(4/3) (m^{\ast}/m_{0}) (1/\sqrt{2 {m^{\ast}}^{3} E_{g}}) (\eta/
\sqrt{1- \eta/3})$\ \cite{aronov} with $m_{0}$ denoting the free electron mass,
 $E_{g}$ being the band gap and $\eta= \Delta/(\Delta + E_{g})$. $\Delta$ is the SOC of the valence band. Moreover, a
footprint of the Coulomb scattering on the spin R/D is predicted at
low temperature when the electron density is not too high and the
impurity density is low.\cite{zhou}  L\"{u} {\em 
et al}. applied the KSBEs to study
the heavy and light holes in (001) GaAs QWs\cite{lu} where the SOC is due
to the Rashba term\cite{winkler} and is very strong so that the system is in
the weak scattering limit. Therefore the single-particle formula
Eq. (1) fails and only the KSBE approach is applicable. It is shown
that in the weak scattering limit, adding a new scattering, such as
the Coulomb scattering provides an additional spin R/D channel so that
the spin R/D time is shorter.\cite{lu} This is in opposite to the case of
the strong scattering limit.\cite{weng2, lu} Finally, it is pointed out
in Ref.\ \onlinecite{lu2} that due to the strong Coulomb scattering, $T_{1}=
T_{2}= T_{2}^{\ast}$ is valid over a very
wide temperature and density regime. 


\section{Spin diffusion/transport}

By solving the KSBEs together with the Poisson equation
self-consistently, one is able to obtain all the transport properties
such as the mobility, charge diffusion length and spin
diffusion/injection length without any fitting parameter. It was first
pointed out by Weng and Wu\cite{weng3} that the drift-diffusion equation
approach is inadequate in accounting for the spin
diffusion/transport. It is important to include the off-diagonal term
$\rho_{{\mathbf k} \uparrow \downarrow}$
in studying the spin diffusion/transport. With this term, electron
spin precesses along the diffusion and therefore ${\mathbf k} \cdot
{\mathbf \nabla}_{{\mathbf r}} \rho_{{\mathbf k}} ({\mathbf r}, t)$  in the diffusion
term offers an additional inhomogeneous broadening. With this
additional inhomogeneous broadening, any scattering, including the
Coulomb scattering, can cause an irreversible spin R/D.\cite{weng3} Unlike
the spin precession in the time domain where the inhomogeneous
broadening is determined by ${\mathbf h} ({\mathbf k})$, here it is
determined by $|g \mu_{B} {\mathbf B} + {\mathbf h} ({\mathbf k})|/ k_{x}$
 provided the
diffusion is along the $x$-axis. Therefore, even in the absence of the
DP term, the magnetic field alone can provide an inhomogeneous
broadening.\cite{weng3} Moreover, it is first pointed out that a spin pulse
can oscillate along the diffusion in the absence of the magnetic field
at very high temperature.\cite{weng4} Detailed study is performed later on
this effect.\cite{weng5,jiang} This oscillation was later realized
experimentally by Crooker and Smith\cite{ohno} in bulk system at very low
temperature. It is also shown that the Coulomb drag plays a very
important role in the spin diffusion and the electric field can
enhance/suppress the spin diffusion.\cite{jiang} Very recently Cheng and Wu
developed a new numerical scheme to calculate the spin
diffusion/transport\cite{cheng3} with very high accuracy and speed. It is
discovered that due to the scattering, especially the Coulomb
scattering, $T_{2}=T_{2}^{\ast}$ is valid even in the space domain. Moreover, as the
inhomogeneous broadening in spin diffusion is determined by
$|{\mathbf h} ({\mathbf k})|/k_{x}$ in the
absence of magnetic field, the period of the spin oscillations along
the $x$-axis is independent on the electric field\cite{cheng3} which is
different from the spin precession rate in the time domain.\cite{weng2} This
is consistent with the experimental findings by 
Beck {\em et al}.\cite{ohno} Many
properties of the spin diffusion/transport in the steady state are
addressed in detail in Ref.\ \onlinecite{cheng3}.

\section{Summary}

In summary we reviewed our fully microscopic KSBE investigate on spin
R/D in the spin precession and spin diffusion/transport. The
importance of the Coulomb scattering is stressed. This approach allows
one to deal with systems both near and far away from the equilibrium
and in both the strong and weak scattering limits. Moreover, both the
effect of inhomogeneous broadening on the scattering and the counter
effect of the scattering on the inhomogeneous broadening are fully
accounted. 

\begin{acknowledgements}

This work was supported by the Natural Science
  Foundation of China under Grant No. 10574120, the Natural Science
  Foundation of Anhui Province under Grant No. 050460203, the
  Knowledge Innovation Project of Chinese Academy of Sciences and
  SRFDP. One of the authors (M.W.W.) would like to thank National
  Center for Theoretical Sciences and W. C. Chou at National Chiao
  Tung University for hospitality where this work was finalized.

\end{acknowledgements}

\end{document}